\documentclass[aps,prd,twocolumn,showpacs]{revtex4}
\begin{document}
\title{Entanglement in Relativistic Quantum Field Theory}
\author{Yu Shi}

\affiliation{Center for Advanced Study, Tsinghua University,
Beijing 100084, China\\
Department of Physics, University of Illinois at Urbana-Champaign,
Urbana, IL 61891, USA}

\begin{abstract}
I present some general ideas about quantum entanglement in
relativistic quantum field theory, especially entanglement in the
physical vacuum. Here, entanglement is defined between different
single particle states (or modes), parameterized either by
energy-momentum together with internal degrees of freedom, or by
spacetime coordinate together with the component index in the case
of a vector or spinor field. In this approach, the notion of
entanglement between different spacetime points can be
established. Some entanglement properties are obtained as
constraints from symmetries, e.g., under Lorentz transformation,
space inversion, time reverse and charge conjugation.
\end{abstract}


[Phys. Rev. D 70, art. no. 105001 (2004)]

 \maketitle

Quantum entanglement is a notion about the structure of a quantum
state of a composite system, referring to its non-factorization in
terms of states of subsystems. It is regarded as an essential
quantum characteristic~\cite{epr,peres0}. Entanglement with
environment is also crucial in decoherence, i.e. the emergence of
classical phenomena in a quantum foundation, and may even be a
possible explanation of superselection rules~\cite{deco,zeh0}.
There has been a lot of activities on various aspects of
entanglement, including some recent works which take into account
relativity~\cite{peres,verch}. Investigations on entanglement in
quantum field theories may provide useful perspectives on field
theory issues. On the other hand, as the framework of fundamental
physics, incorporating relativity, quantum field theory may be
useful in deepening our understanding of entanglement. Besides,
entanglement due to environmental perturbation may also be helpful
in understanding spontaneous symmetry breaking. Most of the
methods in field theory adopt Heisenberg or interaction picture,
and do not need the explicit form of the underlying quantum state
living in an infinite-dimensional Hilbert space. Nevertheless, in
many circumstances, it is still important to know the nature of
the quantum state, most notably the vacuum. In this paper, as an
extension of some previous discussions on non-relativistic quantum
field theories~\cite{shi1}, I present some general ideas about the
nature of entanglement in relativistic quantum field theory, and
constraints from symmetries. Such investigations may offer useful
insights on the structures of the vacua in quantum field theories
on one hand, and on quantum information in relativistic regime on
the other hand.

First, I describe the basic method here of characterizing
entanglement in quantum field theory.  In quantum field theory,
the dynamical variables are field operators (in real spacetime) or
annihilation and creation operators (in energy-momentum space), in
terms of which any observable can be expressed. Spacetime
coordinate  plus the component index in the case of a vector or
spinor field, or energy-momentum plus internal degrees of freedoms
(such as being particle or antiparticle, spin, polarization, etc.)
are merely parameters.  These parameters define the modes in
either the real spacetime or the momentum space, and exactly
provide the labels for the (distinguishable) subsystems, between
which entanglement can be well defined, in the same manner as that
for distinguishable non-relativistic quantum mechanical systems.
In other words, consider the Hilbert space as composed of the
Hilbert spaces for all the modes, parameterized either by the
spacetime or by the momentum, together with whatever other degrees
of freedoms.  Therefore, in momentum space, a mode, parameterized
by the energy-momentum together with internal degrees of freedom,
is entangled with other modes if the quantum state cannot be
factorized as a direct product of the state of this mode and the
rest of the system. Similarly, in real spacetime, a mode
parameterized by the spacetime coordinate is entangled with other
modes if the quantum state cannot be factorized as a direct
product of the state of this mode and the rest of the system. The
basis of the Hilbert space at each specified mode can be
arbitrarily chosen to be a orthonormal set of eigenstates  at this
mode. A convenient, but not necessary, basis of the modes in the
momentum space is the occupation number states, as previously used
in some related investigations~\cite{shi1,zanardi,shi2,enk}. In
the real spacetime, one can use the eigenstates of the local
density $\phi^{\dagger}(x)\phi(x)$, where $\phi(x)$ is the field
operator. The concept of ``local operation'', as used in theories
of entanglement, is generalized to an operation only acting on a
subsystem. In real spacetime, this generalization is consistent
with the usual meaning, but I have naturally incorporated
relativity: one can consider {\em entanglement between different
spacetime points}.

When different fields, i.e. particle species, coexist, these
different fields are clearly distinguishable subsystems, between
which entanglement can be defined.  In some effective or
approximate theories, different fields may be related by an
additional symmetry, e.g. the isospin in nuclear physics,  and
thus can be treated as a single field with an additional degree of
freedom. In a semi-classical setting, entanglement between fields
and charges was discussed previously~\cite{kiefer}.

In a composite system, there is a complex pattern of entanglement,
which is still only partially understood in theories of
entanglement. For simplicity,  here we focus on the bipartite
entanglement between a subsystem and its complementary subsystem,
i.e. the rest of the system.

In the following, we first stay in momentum space until we shift
to real spacetime later on.

It is instructive to start with the simple case of free field
theories. Under canonical quantization, the Hamiltonians can be
written as ${\cal H} = \int d^3k k_0 N_{k}$ for a real scalar
field, where $N_k$ is particle number operator, ${\cal H} = \int
d^3k k_0 (N_k+N^c_k)$ for a complex scalar field,  ${\cal H} =
\int d^3k k_0 \sum_{\sigma} (N_{k,\sigma}+N^c_{k,\sigma})$ for a
vector or spinor field, and ${\cal H} = \int d^3k k_0
\sum_{\sigma=1,2} N_{k,\sigma}$ for the electromagnetic field in
Coulomb gauge quantization.  Here the superscript $c$ represents
charge conjugation or antiparticles, $\sigma$ represents spin or
polarization. In the vacuum state, the occupation number of each
mode, labelled by four-momentum $k$, together with being particle
or antiparticle, spin or polarization, is $0$.  Thus in momentum
space, mode entanglement trivially vanishes in the vacuum state of
a free field theory. Moreover, in a Hamiltonian eigenstate with a
definite number of particles in a mode, the state of this mode can
be factorized out, and thus there is no entanglement between this
mode and other modes. However, because of degeneracy, e.g., the
four-momentum $(k_0, \mathbf{k})$ and spin may be different even
though $k_0$ is the same,  a Hamiltonian eigenstate is not
necessarily non-entangled.

There is subtlety in  Lorenz gauge quantization of electromagnetic
field. The Hamiltonian is ${\cal H} = \int d^3k k_0 (\sum_{\sigma
=1}^3 N_{k,\lambda}-N_{k,0}) = \int d^3k k_0 \sum_{\sigma=1,2}
N_{k,\sigma}$, under the Gupta-Bleuler condition
$(a_{k,0}-a_{k,3})|\Psi\rangle$ for any physical state.
Consequently the nature of physical modes, with $\sigma =1,2$, is
the same as in Coulomb gauge, as it should be. It can also been
seen that the unphysical modes are entangled with each other,
while they are separated from the physical modes, as they should
be; if they were entangled with physical modes, the physical modes
would unreasonably live in a mixed state.

In general, presence of interaction, including gauge coupling, may
induce nonvanishing entanglement, as in interacting field theories
and even in pure a non-abelian gauge field, where there is
self-interaction.

Now are given some constraints on the nature of entanglement,
imposed by symmetry properties.

A symmetry transformation ${\cal T}$ induces a unitary
transformation $U({\cal T})$ on the quantum state $|\Psi\rangle$
of the system, i.e. \begin{equation} |\Psi\rangle \rightarrow
|\Psi'\rangle = U({\cal T})|\Psi\rangle. \label{psi}
\end{equation}

Under symmetry transformation, the labels of the modes are also
transformed, as given by the standard transformations of the
single particle basis states. This is just a relabel, no matter
whether the quantum state of the system  is invariant under the
transformation. There are two cases, as expounded below.

The word ``mode'' is somewhat ambiguous. Here it really means the
single particle basis state. For example, a single particle state
with momentum {\bf p} and spin $\sigma$ is $|{\bf p},\sigma)
\equiv a^{\dagger}_{{\bf p},\sigma}|0\rangle$, while a
one-particle state at coordinate $x$, with vector or spinor
component $l$, is $|x,l) \equiv
\phi_l^{\dagger}(x)|0\rangle$~\cite{round}. The transformation of
the annihilation operator or field operator can be obtained from
the transformation of the corresponding single particle state.
From the definition of creation operator and the fact that
$|0\rangle$ is always invariant, one knows that the creation
operator transforms in the same way as the single particle
state~\cite{weinberg,zeh2}.

The Case I of mode transformation, under a symmetry
transformation, is that a mode $\alpha$ is relabelled as mode
$\alpha'$ existing in the same basis. The single particle state
$|\alpha')$ is related to single particle state $|\alpha)$ as
$$
|\alpha) \rightarrow  U({\cal T})|\alpha) = |\alpha'),$$ which is
equivalent to
$$a_{\alpha}^{\dagger} \rightarrow
U({\cal T}) a_{\alpha}^{\dagger}U^{\dagger}({\cal T}) =
a_{\alpha'}^{\dagger},$$ where $a_{\alpha}^{\dagger}|0\rangle
\equiv |\alpha)$ while $a_{\alpha'}^{\dagger}|0\rangle \equiv
|\alpha')$.

Such a transformation means that in the mode expansion of the
quantum state $|\Psi\rangle$, the label $\alpha$ is changed to
$\alpha'$. The state $|\Psi\rangle$ itself is changed to
$|\Psi'\rangle$ as given in (\ref{psi}).

Consider, in a quantum state $|\Psi\rangle$,  the entanglement
between mode $\alpha$ and its complementary subsystem, denoted as
$E_{|\Psi\rangle}(\alpha)$. Clearly,
$$E_{|\Psi\rangle} (\alpha)
\equiv E_{|\Psi\rangle'}(\alpha').$$

Now, {\em if the state $|\Psi\rangle$ respects a symmetry}, then
$|\Psi'\rangle = |\Psi\rangle$. Such is the case of the vacuum of
a quantum field theory with a symmetry. Then the nature of
entanglement, as a function of the state, should also be invariant
under this symmetry, i.e. $E_{|\Psi\rangle} (\alpha) =
E_{|\Psi\rangle'}(\alpha)$. Thus
\begin{equation}
E_{|\Psi\rangle}(\alpha)=E_{|\Psi\rangle}(\alpha'). \end{equation}

This equality is true no matter what is the specific measure of
$E_{|\Psi\rangle}(\alpha)$. But it can be confirmed for specific
entanglement measures. It is now well known that for a pure state,
the entanglement between a subsystem and the rest of the system is
quantified as the von Neumann entropy of the reduced density
matrix of either subsystem~\cite{bennett}.  Thus $E(\alpha)$ can
be quantified as
$$E(\alpha)=-Tr\rho([\alpha])\ln \rho([\alpha]),$$ where
the trace is over the Hilbert space of all the system excluding
mode $\alpha$,
$$\rho[\alpha] = \sum_N {_{\alpha}\langle} N |\Psi\rangle
\langle \Psi|N\rangle_{\alpha},$$ is the reduced density matrix of
the subsystem complementary with $\alpha$, obtained by tracing
over the Hilbert space at $\alpha$. $|N\rangle_{\alpha} \equiv
(1/\sqrt{N!}){a_{\alpha}^{\dagger}}^N|0\rangle_{\alpha}$ is the
particle number state at $\alpha$. Indeed, under the symmetry
transformation, $|N\rangle_{\alpha} \rightarrow
|N\rangle_{\alpha'} \equiv U({\cal T})|N\rangle_{\alpha}$.
Therefore if $|\Psi\rangle = U({\cal T})|\Psi\rangle$, then
$\rho([\alpha])= \rho([\alpha'])$, and thus
$E_{|\Psi\rangle}(\alpha)=E_{|\Psi\rangle}(\alpha')$.

To summarize for this point, {\em if the state is invariant under
a symmetry transformation, then in the same state, for any two
modes that can be transformed into each other under a symmetry
transformation, they have the same amount of entanglement with the
corresponding complementary subsystems}. The statement is of
course also true if one mode is replaced as a set of modes.

Symmetries of space inversion $P$, time reverse $T$ and charge
conjugation $C$ belong to this case. So does the invariance of a
scalar field under Lorentz transformation $\Lambda$ (translation
has no effect on momentum, so only homogeneous Lorentz
transformation needs to be considered here).

Any vacuum state must be invariant under Lorentz transformation
and $CPT$. This has consequences on the entanglement properties,
as given below.

For a scalar field, the single particle state $|p)$ is transformed
as $U(\Lambda)|p) = \sqrt{(\Lambda p)^0/p^0} |\Lambda p)$ under a
homogeneous Lorentz transformation $\Lambda$, as $P|p) = \eta
|{\cal P}p)$ under space inversion, as $T|p) = \zeta |{\cal P}p)$
under time reversal,  and as $C|p,n) = \xi_n |p,n^c)$ under charge
conjugation, where $n$ denotes the particle species, ${\cal P}p=
(p_0,-\mathbf{p})$, $\eta$, $\zeta$ and $\xi_n$  are phase factors
only dependent on particle species. Phase change of the single
basis particle state does not affect the entanglement between
modes. Thus in a vacuum state, for any mode $p$ of a scalar field,
$E(p)=E(\Lambda p)$ for any $\Lambda$. If it is invariant under
$P$ or $T$, then $E(p)=E({\cal P}p)$. If it is invariant under
$C$, then $E(p,n)=E(p,n^c)$. Consequently $CPT$ theorem implies
that $E(p,n)=E(p,n^c)$ always holds.

Now consider a vector or spinor field. For a massive  field,
$P|p,\sigma) = \eta|{\cal P}p,\sigma)$, $T|p,\sigma) =
\zeta(-1)^{j-\sigma}|{\cal P}p,-\sigma)$, where $j$ is the spin
quantum number, $\sigma$ runs from $j$ to $-j$. For a massless
field, $P|p,\sigma) = \eta_{\sigma}\exp(\mp i\pi\sigma)|{\cal
P}p,-\sigma)$, $T|p,\sigma) = \zeta_{\sigma}\exp(\pm
i\pi\sigma)|{\cal P}p,\sigma)$. The notations are
standard~\cite{weinberg}. The single particle phase factors have
no effect on entanglement. Thus for massive field modes, $P$
symmetry implies $E(p,\sigma)=E({\cal P}p,\sigma)$, while $T$
symmetry implies $E(p,\sigma)=E({\cal P}p,-\sigma)$. For massless
field modes, $P$ symmetry implies $E(p,\sigma)=E({\cal
P}p,-\sigma)$, while $T$ symmetry implies $E(p,\sigma)=E({\cal
P}p,\sigma)$. Note the difference between massive and massless
fields.  For both massless and massive fields,
$C|p,\sigma,n)=\xi_n|p,\sigma,n^c)$, hence $C$ symmetry means
$E(p,\sigma,n)=E(p,\sigma,n^c)$. $CPT$ theorem implies that
$E(p,\sigma, n)=E(p,-\sigma,n^c)$ always holds for both massless
and massive fields.

Lorentz transformation for a vector or spinor field, which mixes
modes with different spins, belong to a different case. Let's
refer to it as Case II, in which a mode is transformed to a
superposition of more than one mode in the single particle basis
considered, i.e.
$$|\alpha) \rightarrow U({\cal T})|\alpha) =
\sum_i\gamma_i|\alpha_i),$$ where $\gamma_i$ represents
coefficients. In other words,$$a_{\alpha}^{\dagger} \rightarrow
Ua_{\alpha}^{\dagger}U^{-1} =\sum_i \gamma_i
a^{\dagger}_{{\alpha}_i}.$$

In this case, the occupation-number states at mode $\alpha$
transform as $|0\rangle_{\alpha} \rightarrow \prod_i
|0\rangle_{\alpha_i}$ and $|N\rangle_{\alpha} \rightarrow
U|N\rangle_{\alpha} = \frac{1}{\sqrt{N!}} (\sum_i \gamma_i
a^{\dagger}_{{\alpha}_i})^N \prod_i |0\rangle_{\alpha_i}$.

Therefore, if the quantum state $|\Psi\rangle$ respects the
symmetry, i.e. $U|\Psi\rangle =|\Psi\rangle$, then the reduced
density matrix $\rho([\alpha])$ must satisfy $$\rho([\alpha])=
\sum_N \frac{1}{N!} \langle {\bar{0}}|(\sum_i \gamma_i^*
a_{{\alpha}_i})^N|\Psi\rangle \langle \Psi|\sum_i \gamma_i
a_{{\alpha}_i}^{\dagger})^N |{\bar{0}}\rangle,$$ where
$|\bar{0}\rangle \equiv \prod_i |0\rangle_{\alpha_i}$.

Now gauge transformation is considered, which  changes the phase
of the field operator, accompanied by the transformation of the
gauge potential. Non-abelian gauge transformation involves a local
rotation between different components of the spinor or vector
field. Consider a field operator $\phi(x)$, be it scalar, vector
or spinor. It  is gauge transformed as $\phi(x) \rightarrow
\phi'(x)= S(x)\phi(x)$. Consequently,  a creation operator
$a_{p,\sigma}^{\dagger}$, obtained from the momentum-spin mode
expansion of $\phi(x)$, is transformed a new mode creation
operator ${a'}^{\dagger}_{p,\sigma}$, obtained from the mode
expansion of $\phi'(x)$. $a_{p,\sigma}$ and
${a'}^{\dagger}_{p,\sigma}$, however, act on the same mode
$(p,\sigma)$. The entanglement $E(p,\sigma)$ is thus transformed
to itself. It is consistent, though no particular constraint on
entanglement is obtained from this simple consideration.

Now switch to real spacetime, in which there exists entanglement
even in the vacuum of a free field,  as simply seen by
transforming the creation operators in momentum space to field
operators in real spacetime. This seems consistent with the early
result about violation of Bell inequalities in vacuum
states~\cite{summers}. Very recently, Calabrese and Cardy made
some calculations on  positional entanglement in 1+1 dimensional
field theory~\cite{cardy}.

My discussion here is fully relativistic; the subsystems are
spacetime points.

When the quantum states are represented in the real spacetime,
there is a degree of freedom in addition to the spacetime
coordinate, namely, the  component index of the irreducible
representation of the homogeneous Lorentz group, which defines the
field operator. However, one need not explicitly consider the
vector or spinor components, rather, we can use the whole vector
or spinor, since in field theories, the Lagrangians can be written
in terms of the whole vector or spinor. Of course, one also needs
to consider all different fields in the system. In this way, one
can obtain the total entanglement between different spacetime
points.

One may use eigenstates of an hermitian operator as the basis for
the Hilbert space at $x$. For example, such an hermitian operator
can be chosen to be the local density $D(x)$, which is defined to
be $\phi^{\dagger}(x)\phi(x)$ for a scalar field $\phi(x)$,
$v^{\dagger}(x)v(x)$ for a vector field $v(x)$, and
$\psi(x)^{\dagger}\psi(x)$ for a spinor field $\psi(x)$.

It can be checked that for each of these fields, $D(x)$ is a
scalar under a Lorentz transformation $x\rightarrow x'=\Lambda x +
l$, i.e. $D(x)=D(x')$. The Lorentz invariance of the state
$|\Psi\rangle$ means that $E(x)=E(y)$, where $x$ and $y$ are any
two spacetime points that can be connected by a Lorentz
transformation.  $P$ transforms $D(x)$ to $D({\cal P}x)$, $T$
transforms $D(x)$ to $D({-\cal P}x)$. Hence $P$ symmetry implies
$E(x)=E({\cal P}x)$, $T$ symmetry implied $E(x)=E(-{\cal P}x)$.
$C$ transformation transforms $D(x)$ to itself, so no special
constraint is given by $C$ symmetry. Therefore $CPT$ symmetry
implies that $E(x)=E(-x)$ always holds.

Because these symmetry transformations of the entanglement are,
respectively, the same for different fields, they remain the same
when different fields coexist.

A global gauge transformation is merely a constant phase factor,
so trivially has no effect on entanglement. The local gauge
transformation only depends on the local spacetime, therefore also
does not have any effect on the entanglement between different
spacetime points. In fact, the underlying quantum state of the
field theory remains the same under any gauge transformation.

I stress that the entanglement between different spacetime points,
obtained by tracing over the spinor or vector components and over
different fields, is an intrinsic physical property of the system
in consideration. Although a particular momentum-spin mode defined
by a free single particle basis state may not be directly
measurable because of renormalization, the entanglement between
spacetime regions is directly measurable in principle.

This interesting point can be illustrated by using the well-known
entanglement~\cite{enthaw,zeh0,pad,zeh} in Unruh-Hawking
radiation~\cite{unruh,hawking}. As shown by Unruh~\cite{unruh},
the Minkowski vacuum can be expressed in terms of Rindler modes,
which are those in the accelerating frame, as
\begin{equation}
\begin{array}{lll}
|vac\rangle & \propto&  \prod_{\omega,\mathbf{k}}
\exp[e^{-2\pi\omega}
a^{\dagger}_{1,\omega,\mathbf{k}}a^{\dagger}_{2,\omega,\mathbf{k}}]|0\rangle_R
\\&= & \prod_{\omega,\mathbf{k}}\sum_n e^{-2\pi
n\omega}|n\rangle_{1,\omega,\mathbf{k}}
|n\rangle_{2,\omega,\mathbf{k}},
\end{array}\label{rindler}
\end{equation}
where the subscripts $1$ and $2$ represent the two halves of the
Rindler space separated by the horizon. They must appear, together
with the energy-momentum, as the subscripts,  because for each
half of the Rindler space, there is a set of momentum-mode
functions, which vanish in the other part of the Rindler space. So
the momentum-mode functions in both halves are needed to make a
complete set.  It can be seen that the entanglement between modes
$(1, \omega,\mathbf{k})$ and  $(2, \omega,\mathbf{k})$, equal to
the entanglement between each of them and the rest of the whole
system, is $S_{\omega,\mathbf{k}}=-\sum_n p_{\omega}(n)\ln
p_{\omega}(n)$, where $p_{\omega}(n) = e^{-4\pi n\omega}/\sum_n
e^{-4\pi n\omega}$. Similarly, in the exterior of a Schwarzschild
black hole, in terms of the modes on the two sides of the event
horizon, the vacuum state is given in Eq.~(\ref{rindler}) with
$\pi$ replaced as $2\pi M$, where $M$ is the mass of the black
hole. With this replacement,  the nature of entanglement is the
same as that for Minkowski vacuum in terms of Rindler modes.

One can obtain the total entanglement between the two halves of
Rindler space, or the entanglement across the event horizon of a
black hole, as $\sum_{\omega,\mathbf{k}} S_{\omega,\mathbf{k}}$.
This is the entanglement between two parts of the spacetime, a key
concept stressed in this paper. The result is independent of the
choice of the momentum-mode functions during the calculation.

To summarize, I present some general ideas concerning field
theoretic quantum entanglement, and especially its use in
characterizing quantum properties of vacuum,  a key issue in
fundamental physics. Field theoretic entanglement can be defined
in momentum space and in real spacetime, with the
(distinguishable) subsystems parameterized either the
energy-momentum plus internal degrees of freedom, or by the
spacetime coordinate plus the component index for a vector or
spinor field, respectively. With this definition, the ideas from
the theories of entanglement can be applied. I give some symmetry
properties concerning the entanglement in quantum field theory, in
momentum space and in real spacetime, respectively. I discussed
the invariance properties of entanglement when the quantum state
respects symmetries. A noteworthy notion is the entanglement
between different spacetime points, which is an intrinsic physical
property and is measurable in principle. This notion is
illustrated in terms of the entanglement between the two halves of
the Rindler space or across the event horizon of a black hole.

I am very grateful to Professor Michael Stone, Professor Yongshi
Wu  and Professor H. Dieter Zeh for useful discussions.


\begin{thebibliography}{99}
\bibitem{round} Here I use a parenthesis in the
ket notation of a single particle state, to avoid confusion with
the ``second-quantized'' states, which are denoted as the usual
kets.
\end{thebibliography}
\end{document}